# Regulatory Options and Technical Challenges for the 5.9 GHz Spectrum: Survey and Analysis

Junsung Choi, Vuk Marojevic, *Senior Member, IEEE*, Mina Labib,
Siddharth Kabra, Jayanthi Rao, Sushanta Das, *Senior Member, IEEE*,
Jeffery H. Reed, *Fellow, IEEE,* and Carl B. Dietrich, *Senior Member, IEEE*

*Abstract*—In 1999, the Federal Communications Commission (FCC) allocated 75 MHz in the 5.9 GHz ITS Band (5850-5925 MHz) for use by Dedicated Short Range Communications (DSRC) to facilitate information transfer between equipped vehicles and roadside systems. This allocation for DSRC in the ITS band has been a co-primary allocation while the band is shared with the Fixed Satellite Service (FSS), fixed microwave service, amateur radio services and other Federal users authorized by the National Telecommunications and Information Administration (NTIA). In recent time, Cellular V2X (C-V2X), introduced in 3GPP Release 14 LTE standard, has received significant attention due to its perceived ability to deliver superior performance with respect to vehicular safety applications. There is a strong momentum in the industry for C-V2X to be considered as a viable alternative to DSRC and accordingly, to operate in the ITS spectrum. In another recent notice, the FCC is soliciting input for a proposed rulemaking to open up more bandwidth for unlicensed Wi-Fi devices, mainly based on the 802.11ac standard. The FCC plans to work with the Department of Transportation (DoT), and the automotive and communications industries to evaluate potential sharing techniques in the ITS band between DSRC and Wi-Fi devices. This paper analyzes the expected scenarios that might emerge from FCC and the National Highway Traffic Safety Administration (NHTSA) regulation options and identifies the technical challenge associated with each scenario. We also provide a literature survey and find that many of resulting technical challenges remain open research problems that need to be addressed. We conclude that the most challenging issue is related to the interoperability between DSRC and C-V2X in the 5.9 GHz band and the detection and avoidance of harmful adjacent and co-channel interference.

## I. Introduction

Dedicated Short Range Communication (DSRC) has been the dominant protocol recommended for vehicular communication such as Vehicle-to-Vehicle (V2V) and Vehicle-to-Infrastructure (V2I). The Federal Communications Commission (FCC), in its Report and Order FCC-03-324 [1], allocated 75 MHz in the 5.9 GHz band for vehicular communications. DSRC is using these 75 MHz with seven different channels; each is 10 MHz wide as per the recommendation of Intelligent Transportation Systems (ITS) with 5 MHz reserved band before these channels. More recently, the FCC issued a Notice of Proposed Rulemaking (NPRM) regarding the potential use of the 5.9 GHz band for Unlicensed National Information Infrastructure (U-NII) devices. Also, the FCC is considering sharing the 5.85 - 5.925 GHz band between DSRC and U-NII devices, according to the FCC Docket ET 13-49 [2]. The primary unlicensed devices considered in the FCC NPRM use a signal based on IEEE 802.11ac that operate in the U-NII-4 band.

The channel allocations being considered for DSRC and U-NII-4 are shown in Fig. 1. In the FCC NPRM, two interference mitigation approaches are presented: detect and vacate (DAV) and re-channelization. DAV represents no changes to DSRC. It requires unlicensed devices to avoid interfering with the DSRC signal by detecting the DSRC signal up to channel 178. Re-channelization is an allocation process where safety-related DSRC applications use the upper 30 MHz (channels 180, 182, and 184) while non-safety-related DSRC and U-NII devices share the lower 45 MHz (channels 172, 174, 176, and 178).

New developments for Vehicle-to-Everything (V2X), are expected over the next years. One prmising apporach is to provide services through the long-term-evolution (LTE) based

This work was supported by Ford Motor Company through the University Alliance Program.
S. Das is the Ford PI for this project.
J. Choi, M. Labib, S. Kabra, J. H. Reed, and C. B. Dietrich are with the Wireless@VT, Bradley Department of Electrical and Computer Engineering, Virginia Tech, Blacksburg, VA 24061 USA (e-mail: choijs89@vt.edu; mlabib@vt.edu; skabra27@vt.edu; reedjh@vt.edu; cdietric@vt.edu)
V. Marojevic is with the Electrical and Computer Engineering Department of Mississippi State University, Miss. State, MS 39762, USA (email: vuk.marojevic@msstate.edu)
J. Rao and S. Das are with Ford Motor Company, Dearborn, MI 48126, USA (email: jrao1@ford.com; sdas30@ford.com)

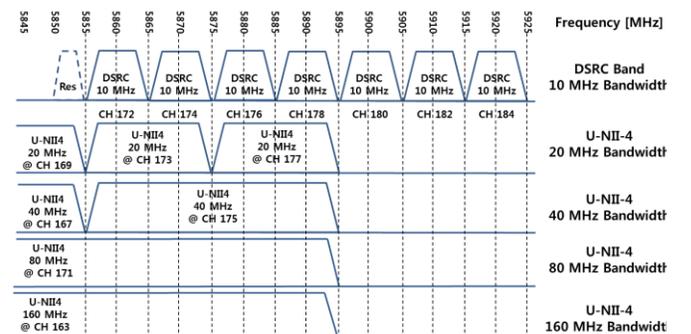

Fig. 1. Proposed U-NII4 channels and existing DSRC channels allocation

TABLE I
ITS BAND REGULATORY DECISION PLANNING

| | | | One Safety Technology in Band | | | | | | | | Two or More Safety Technology in Band | | | | | | | |
|---|---|---|---|---|---|---|---|---|---|---|---|---|---|---|---|---|---|---|
| | | | IEEE 802.11p based DSRC | | | | 3GPP R14 based Cellular-V2X | | | | Co-Channel Coexistence | | | | Adjacent Channel Coexistence | | | |
| | | | No R-CH | | R-CH | | No R-CH | | R-CH | | No R-CH | | R-CH | | No R-CH | | R-CH | |
| | | | No Wi-Fi | Wi-Fi DAV | No Wi-Fi | Wi-Fi DAV | No Wi-Fi | Wi-Fi DAV | No Wi-Fi | Wi-Fi DAV | No Wi-Fi | Wi-Fi DAV | No Wi-Fi | Wi-Fi DAV | No Wi-Fi | Wi-Fi DAV | No Wi-Fi | Wi-Fi DAV |
| NHTSA ITS Spectrum Regulatory Decision Criteria | | DSRC Only | c | a | b | a, b | | | | | | | | | | | | |
| | | C-V2X Only | | | | | f | d | e | d, e | | | | | | | | |
| | Coexistence | Interoperable | | | | | | | | | c, f, h, i | a, d, h, i | b, e, h, i | a, b, d, e, h, i | | | | |
| | | Non Interoperable | | | | | | | | | c, f, h | a, d, h | b, e, h | a, b, d, e, h | c, f, g | a, d, g | b, e, g | a, b, d, e, g |
| | | No Reg. | j | | | | | | | | | | | | | | | |

TABLE II
LIST OF TECHNICAL CHALLENGES

| Alphabet Code from Table I. | Description | Alphabet Code from Table I. | Description |
|---|---|---|---|
| a | Ability of Wi-Fi to detect DSRC signals to vacate | f | Co-channel interference between Wi-Fi and C-V2X |
| b | Adjacent interference to DSRC: out-of-band rejection of DSRC | g | Analyzing adjacent interference from DSRC to C-V2X and from C-V2X to DSRC. Depending on which system will use the lower frequency band, adjacent interference from Wi-Fi need to be analyzed |
| c | Co-channel interference between Wi-Fi and DSRC | h | How the co-channel coexistence between DSRC and C-V2X will affect the performance to each other |
| d | Ability of Wi-Fi to detect C-V2X signals to vacate | i | Interoperability between DSRC and C-V2X |
| e | Adjacent interference to C-V2X: out-of-band rejection of C-V2X | j | The "No Regulation" will inherit all technical issues in the column above it |

system known as Cellular-Vehicle-to-Everything (C-V2X) [3]. Recently, 3GPP presented the sidelink interface and the LTE cellular interface in Release 14 with fulfilling the requirements of V2X services [4]. Important changes are expected for 5G in upcoming 3GPP releases.

As Wi-Fi, LTE, and DSRC are competing for the 5.9 GHz spectrum, several technical and research challenges arise and are reviewed in this paper.

DSRC is the dominant protocol for 5.9 GHz; however, allowing C-V2X, Wi-Fi, and other unlicensed devices to access it is also being considered. Within the possible hypotheses about the directions provided by the FCC and the National Highway Traffic Safety Addminstration (NHTSA), we generate possible combinations of regulatory decisions and the associated technical challenges. Moreover, we present a comperhensive technical assessment that is related to the potential technical challenges for those regulations.

## II. POSSIBLE REGULATION SCENARIOS

FCC and NHTSA play major roles for the adoption of vehicular communication standards. Their decisions may or may not conflict; however, it is obvious that C-V2X or DSRC or both will be using the 5.9 GHz. Based on the announced NPRM, FCC's current position is to use UNII-4 and DSRC with Re-channelization.

FCC is responsible for spectrum regulations and allocating spectrum for use by specific technologies. NHTSA is interested in ensuring road safety and supporting technologies that can satisfy the established road safety requirements. Within these boundaries, we foresee different scenarios.

Possible FCC actions are:
- Number of vehicular safety communication technologies in band: 1) one technology; or 2) two technologies
- If one safety technology in band: 1) DSRC Only; or 2) C-V2X Only
- If two or more safety technologies in band: 1) co-channel coexistence; or 2) adjacent channel coexistence
- Re-channelization: 1) Re-channelize, safety-related DSRC using the upper 30 MHz and unlicensed devices using the lower 45 MHz; or 2) Do not re-channelize
- Wi-Fi sharing: 1) Allowed through DAV; 2) Not allowed
- Other unlicensed technologies, most prominently LTE-U, LAA, and MuLTEfire: 1) Allowed; 2) Not allowed

Re-channelization and Wi-Fi sharing by DAV are considered as two different interference mitigation approaches. However, they can be used at the same time: using safety-related DSRC to re-channelize in the upper 30 MHz and allocate non-safety-related DSRC in the lower 45 MHz, allowing Wi-Fi sharing through DAV.

We identify the possible NHTSA's regulations as:
- Only DSRC operating in the ITS band
- Only C-V2X operating in the ITS band
- DSRC & C-V2X co-operating in the ITS band
  - Interoperable
  - Non-Interoperable
- No regulations

The scenario of coexistence can be considered as interoperable or non-interoperable. Interoperable allows DSRC and C-V2X to communicate; one device to communicate between DSRC and C-V2X.

With the combinations of identified FCC and NHTSA regulation scenarios, we generate the scenario chart shown in Table I. The scenarios that did not filled out with an alphabet code are not possible. For each combination of scenarios, certain technical challenges are inherent. The technical challenges are shown as alphabet code in Table I. The details of these technical challenges are listed in Table II.

## III. TECHNICAL SURVEY

By categorizing the technical challenges in Table II, we surveyed papers and categorized them into the related technical challenge topics. Aligned with the current plan of DSRC using the entire 5.9 GHz band alone, we found the greatest number of papers. The fewest studies are devoted to the technical challenges of coexistence between DSRC and C-V2X. Therefore, in this section, we identify the technology gaps that have not been addressed in the technical literature. Table III summarizes the technical challenges that are addressed in open literature and their key findings.

*A. DSRC Only*

Reference [5] evaluates the effects of adjacent channel interference in multi-channel vehicular networks. In the model setup, a target node is observing in Service Channel (SCH) 4 and various numbers of nodes transmit on SCH3 to cause adjacent channel interference on the target node. Their study shows that a node tuning into a channel with a low transmission power, so to mitigate adjacent channel interference effects would preserve the communication quality to some extent. The study also concludes that despite the blocking, channel access delay may be reduced and transmissions may be less prone to collisions.

The authors in [6] analyze the effects of adjacent channel interference levels, channel access delay and packet loss in the multi-channel vehicular networks using an adjacent channel interference model through simulation that is able to control time, space, and frequency parameters for mobile nodes. They explore two scenarios: 1) vehicular nodes are arranged in a square and adjacent channel interference effects are measured in the center; 2) 60 cars are exponentially distributed over 3-lane highway. Through simulations, the effect of adjacent channel interference is found to be significant for transmission power settings of 20 dBm and when the involved nodes are at a distance lower than 7 m. The results show that increased packet losses is the more evident effect of adjacent channel interference in mixed co-channel and adjacent channel interference scenarios.

The impact of adjacent channel interference on the DSRC Control Channel (CCH) due to communication in adjacent SCH channel is evaluated in [7]. The setup considers two devices communicating on CCH and two on SCH. Two nodes, one on each channel, are kept close together while the other two are farther away. Power levels of 5, 10, 15, 20, and 33 dBm and 5-500 m separations are used. The authors observe that 1) a transmit power of 33 dBm is not applicable because the adjacent channel will be found busy by the nearby node; 2) it is not applicable to run both channels with a similar transmit power due to the reduction of the communication range to 100 m on both channels; 3) the best power difference seems to be 10 dB where the best communication range on the CCH and SCH can be achieved.

The impact of on-board 802.11a Wi-Fi device communication with an electronic toll collector on the vehicle's DSRC communication performance is evaluated in [8]. The tests were done with 3 lanes and 100 cars in each lane. Two traffic cases are considered: 1) normal traffic case, in which cars move at 54 km/h and the average inter-car distance is 10 m; 2) traffic jam case, which assumes that cars are moving at 3.6 km/h and the average inter-car distance is 3 m. Also, the transmit power sensitivity is considered by 1) varying the percentage of cars using 802.11a; 2) varying the transmit power of 802.11a; and 3) varying transmit power of the 802.11p RSU. The authors of [8] conclude that the effect of 802.11a interference cannot be eliminated just by increasing the 802.11p transmit power and need further mechanisms to make DSRC more reliable and rugged.

Reference [9] analyzes the coexistence between Wi-Fi and DSRC by analyzing the physical layer challenges and the MAC layer challenges for the two systems. At short distances between DSRC transmitter and receiver there is no significant coexistence issues. For long range DSRC communications, there is high DSRC packet loss due to interference from Wi-Fi, but long distances may not be as critical for safety related DSRC applications. At medium distances, Wi-Fi at outdoor can coexist better than Wi-Fi at indoor, the latter creating non-negligible DSRC packet loss, which can be problematic for safety applications. The results show that even with DAV, Wi-Fi at indoor can cause interference to DSRC, and it is recommended to reduce the Wi-Fi transmit power. Also, the results show that DAV provides better coexistence mechanisms for DSRC, and is a recommended technique if Wi-Fi and DSRC share the same band.

We suggest that additional technical analysis be performed for DSRC Only (F3) regulatory scenario:
- Evaluation of adjacent channel and co-channel Wi-Fi interference effects
- Technical improvement of Wi-Fi to detect DSRC signal for an advanced DAV algorithm

*B. C-V2X Only*

Reference [10] analyzes the effect of the aggregate adjacent channel interference generated from LTE small-cells to a user in a macro-cell. The authors propose an interference approximation model for the interference generated from small-cell to a device connected to the macro-cell as a weighed-sum of lognormal-based distributions. They find that if small-cell density increases, the outage probability of victim users increases and for the same small-cell density, as the distance between victim user and macro-cell base station (BS) increases, the outage probability increases.

The authors of [11] propose a modified OFDM-based scheme for V2X communication to improve performance and robustness of vehicular communication against fast fading, which will happen when vehicles travel at high speed. The authors analyze the effect of the number of subcarriers on the date rate for different relative speeds and conclude that when the data rate increases from 1 Mbps to 10 Mbps under a relative speed of 200 km/h, a larger number of subcarriers is preferred to satisfy the link performance requirements, especially to prevent frequency selective fading.

The authors of [12] evaluate the performance of V2I communication in a freeway scenario in which the coverage is provided by LTE-A. They use an LTE system simulation platform for which the system throughput performance and signal-to-interference-plus-noise ratio (SINR) have been rigorously assessed. For the case in which the network is dense and the reliability requirement high, the results indicate there is a need for novel resource allocation and interference mitigation techniques to meet the performance requirements. For the case in which the minimum and maximum distances between vehicles are 200 m and 300m, respectively (which corresponds to around 40 vehicles connected to each RSU), the result shows that about 50% of vehicles can achieve an SINR of 15 dB and cell edge vehicles (5% from CDFs) can achieve SINR of 2 dB.

The authors of [13] and [14] analyze the performance of LTE-V or C-V2X Mode 4. In [13], the authors compare the performances of DSRC and LTE-V in fast and slow environments and evaluate the LTE-V performance for different modulation schemes. From these studies, the authors observe that LTE-V outperforms DSRC when DSRC is using a low data rate and alternative DSRC due to its improved link budget, the support for redundant transmissions per packet, and different sub-channelization schemes. However, careful configuration of parameters is required for a more efficient use. The authors of [14] analyze the performance of the C-V2X's Semi-Persistent Scheduling (SPS) and evaluate different SPS parameter configurations. They observe that the Packet Delivery Ratio (PDR) improves when increasing the number of available sub-channels or when increasing of resource reservation interval in dense networks.

From our surveys on the research related to C-V2X Only (F4) regulatory scenario, we recommend research on:
- Analysis of C-V2X performance w.r.t. DSRC,
- Quantification of adjacent and co-channel Wi-Fi interference effects on C-V2X
- Technical improvement of Wi-Fi to detect C-V2X signal for an advanced DAV algorithm, and
- Possibilities of improved scheduling and congestion control mechanisms for C-V2X networks

*C. Coexistence between DSRC and C-V2X*

In [15], the 5GAA proposes splitting the lower 30 MHz band between DSRC and C-V2X, where each technology is allocated 10 MHz (5875-5885 MHz, 5895-5905 MHz), and there is 10 MHz in between (5885-5895 MHz), which can be used by either technology through a DAV mechanism. According to 5GAA, the upper 45 MHz allocation will be addressed in the future.

The coexistence between DSRC and C-V2X (F2) and the regulatory implications are open research problems. The technical challenges embrace all challenges mentioned in the previous regulatory scenarios:
- Evaluation of adjacent and co-channel interference effects between DSRC and C-V2X,
- Message scheduling schemes for DSRC and C-V2X,
- Interoperability between DSRC and C-V2X,
- DSRC/C-V2X detection and identification methods,
- Wi-Fi detecting DSRC/C-V2X for advanced DAV algorithm, and
- Advanced channelization, avoidance, and interference mitigation.

## IV. CONCLUSION AND OUTLOOK

We have analyzed the potential regulatory rules that can be made by FCC and NHTSA for the 5.9 GHz spectrum band designated to vehicular communication and discussed possible regulation scenarios. We have derived the main technical challenges for several combinations of scenarios and surveyed related research mapped to the possible scenarios. We have found that some topics have been researched more than the others. Research has addressed the DSRC only scenario, C-V2X only, and to less extent the coexistence of DSRC and C-V2X. We believe that coexistence of DSRC and C-V2X will be the path forward regulation in near future and the technology and market will tell what will be the scenario in the far future. With this in mind, we recommend that interoperability issues

TABLE III
SUMMARY OF TECHNICAL SURVEYS

| Regulatory Scenarios | Contributions |
|---|---|
| DSRC Only | - A node tunes into a channel with a low transmission power to mitigate adjacent channel interference effects would preserve the communication quality at some extent [5].<br>- Interference to nearby nodes gets weaker [5].<br>- Effect of adjacent channel interference should not be neglected for transmission power of 20 dBm at a distance lower than 7 m [6].<br>- Increase of channel access delay due to adjacent channel interference is largely negligible when co-channel interference is also experienced [6].<br>- The best power difference is around 10 dB where the best communication range on the CCH and SCH can be achieved [7].<br>- Effect of 802.11a interference can't be eliminated by increasing 802.11p transmit power [8].<br>- Even with DAV, indoor Wi-Fi can cause interference to DSRC and recommended to reduce Wi-Fi transmit power to avoid creating an interference to DSRC [9]. |
| C-V2X Only | - If small-cell intensity increases, the outage probability of victim user increase [10].<br>- As the distance between victim user and macro-cell BS increases, outage probability increase [10].<br>- When the data rate is increasing from 1 Mbps to 10 Mbps under a relative speed of 200 km/h, the larger number of subcarriers is preferred to satisfy maintain the link, especially to in the presence of frequency selective fading [11].<br>- When the network is dense and the reliability requirements are high, there is a need for novel resource allocation and interference mitigation techniques to meet the performance requirements [12].<br>- LTE-V outperforms DSRC when DSRC is at 6 Mb/s, but DSRC can improve the performance with 18 Mb/s [13].<br>- LTE-V can be an alternative to DSRC due to its improved link budget, the support for redundant transmissions per packet, and sub-channelization schemes; however, a careful configuration of the transmission parameters is required for transmitting more packets per seconds (pps) [13].<br>- PDR improves with the number of available sidelinks sub-channels [14].<br>- PDR increases from 10-25% to 60-85% when the resource reservation interval is increased from 100 to 1000 ms [14]. |
| Coexistence between DSRC and C-V2X | - 5GAA proposes splitting the lower 30 MHz between DSRC and C-V2X and address the upper 45 MHz allocation in the future [15]. |

should be investigated more and account for U-NII4 devices sharing the band with DSRC and C-V2X.

Other than what we presented, another possible coexistence scenario between DSRC and C-V2X would be having DSRC in channel 172 and C-V2X in 184 with a 40 MHz middle portion as a gap. This gap can be explored for coexistence between DSRC, C-V2X, and Wi-Fi. This scenario might not face adjacent channel interference between DSRC and C-V2X; however, the C-V2X performance may be suffer from the effect of adjacent channel interference from U-NII 5 devices operating at 6 GHz.